\documentclass[12pt]{article}

\usepackage{amsfonts,amssymb, graphicx,a4,color}

\definecolor{Red}{cmyk}{0,1,1,0}

\definecolor{verde}{cmyk}{1,0,1,0}

\definecolor{azul}{cmyk}{1,1,0,0}

\let\a=\alpha \let\b=\beta   \let\e=\varepsilon
 \let\g=\gamma      \let\l=\lambda
\let\m=\mu  \let\o=\omega     
 \let\s=\sigma

\global\newcount\numsec\global\newcount\numfor
\gdef\profonditastruttura{\dp\strutbox}
\def\senondefinito#1{\expandafter\ifx\csname#1\endcsname\relax}
\def\SIA #1,#2,#3 {\senondefinito{#1#2}
\expandafter\xdef\csname #1#2\endcsname{#3} \else
\write16{???? il simbolo #2 e' gia' stato definito !!!!} \fi}
\def\etichetta(#1){(\veroparagrafo.\veraformula)
\SIA e,#1,(\veroparagrafo.\veraformula)
 \global\advance\numfor by 1
 \write16{ EQ \equ(#1) ha simbolo #1 }}
\def\etichettaa(#1){(A\veroparagrafo.\veraformula)
 \SIA e,#1,(A\veroparagrafo.\veraformula)
 \global\advance\numfor by 1\write16{ EQ \equ(#1) ha simbolo #1 }}
\def\BOZZA{\def\alato(##1){
 {\vtop to \profonditastruttura{\baselineskip
 \profonditastruttura\vss
 \rlap{\kern-\hsize\kern-1.2truecm{$\scriptstyle##1$}}}}}}
\def\alato(#1){}
\def\veroparagrafo{\number\numsec}\def\veraformula{\number\numfor}
\def\Eq(#1){\eqno{\etichetta(#1)\alato(#1)}}
\def\eq(#1){\etichetta(#1)\alato(#1)}
\def\Eqa(#1){\eqno{\etichettaa(#1)\alato(#1)}}
\def\eqa(#1){\etichettaa(#1)\alato(#1)}
\def\equ(#1){\senondefinito{e#1}$\clubsuit$#1\else\csname e#1\endcsname\fi}

\def\0{\emptyset}

\def\\{\noindent}

\def\vv{\vskip.2cm}

\def\tende#1{\vtop{\ialign{##\crcr\rightarrowfill\crcr
              \noalign{\kern-1pt\nointerlineskip}
              \hskip3.pt${\scriptstyle #1}$\hskip3.pt\crcr}}}
\def\otto{{\kern-1.truept\leftarrow\kern-5.truept\to\kern-1.truept}}

\def\1{\rlap{\mbox{\small\rm 1}}\kern.15em 1}

\def\buildd#1#2{\mathrel{\mathop{\kern 0pt#1}\limits_{#2}}}

\def\Z{\mathbb{Z}}

\BOZZA

\def\be{\begin{equation}}
\def\ee{\end{equation}}

\begin{document}

\title{Diffusive-Ballistic Transition
in Random Walks with Long-Range Self-Repulsion}
\author{Aldo Procacci, R\'emy Sanchis, Benedetto Scoppola}
\maketitle
\begin{abstract}
We prove that a class of  random walks on $\Z^2$
with long-range self-repulsive interactions
have a diffusive-ballistic phase transition.
\end{abstract}

\vv

\vv

\vv\vv\vv
\\\S1. {\it Introduction, model and results}
\vv
\\Random walks  are a simple mathematical tool to
describe polymers,
which are the subject of study of  a rapidly developing area
intersecting physics, chemistry and recently biophysics. In
particular, random walks with self-repulsive long range
interactions are a natural model for polyelectrolytes (see e.g. \cite{dr}).

In this note we propose  a model of random walks with long-range
self-repulsion exhibiting a diffusive-ballistic phase transition. Our arguments stem from
the ideas illustrated in a previous paper \cite{bps} where it was shown that a polymer in two dimensions with a self repelling
interaction of Kac type exhibits a diffusive-ballistic transition if considered on the appropriate scale.
Here we prove that the reasoning and conclusion of [1] can be generalized to infinite range Coulomb-like
interactions (i.e. decaying polynomially) and in this case the phase transition is a genuine one, namely it does not depend
on some length scale as in \cite{bps}.

Self-repulsive random walks with long range interactions are not new in the literature. E.g.
random walk models with some similarity to the model proposed here has been studied in \cite{bmpy},
\cite{mp} and \cite{cpp}. See also \cite{hhs} for rigorous results on a similar model.

The class of random walks in $\Z^2$ that we consider in this paper, denoted ${\cal W}_N$,
is described as a succession of $N$ steps, $\o=(\o_0,\o_1,\dots \o_N)$, with $\o_i\in \Z^2$ and
$|\o_{i+1}-\o_i|=1$, starting at the origin, $\o_0=0$ and weighted according to
\be\label{model}
P_\b(\o)={Z_N}^{-1}\exp\Big[{+\b\sum_{0\leq i<j\leq N}
V_{ij}\cdot(\o_i-\o_j)^2}\Big]
\ee
where $\b>0$ is a parameter which plays the role of the inverse temperature and $Z_N$ is
the normalizing factor given by

\be
{Z}_N(\b)=\sum_{\o\in \,{\cal W}_N}\exp(+\b\sum_{0\leq i<j\leq N} V_{ij}\cdot(\o_i-\o_j)^2).
\ee

The self interaction $V_{ij}$, decreasing with the
difference of the proper time $i$ and $j$ of the random walk, has the following form:
\be\label{poten}
V_{ij}={1\over |i-j|^{\a}}\qquad\qquad 3<\a\le4
\ee
These weights assign greater probability to walks that are ``far apart'' or are stiffer.

Let
\be
\label{quadrado}\left\langle \o_N^2 \right\rangle=\sum_{\o\in\cal{W}_N}\o_N^2P_\b(\o)
\ee
be the mean square end-to-end distance of the walk $\o$ and let
$\g>0$ such that
 $\lim_{N\to \infty}\left\langle \o_N^2 \right\rangle/N^\g$ is finite positive.
A random walk is said to be {\it diffusive} if $\g=1$, {\it
superdiffusive} if $1<\g<2$ and {\it ballistic} if $\g=2$.

We will show that the  self repelling random walk model proposed
in this note is diffusive at sufficiently high temperature and ballistic at
suffciently low temperatures. Our result can be summarized by the following theorem.
\vv
\\{\bf Theorem}. {\it Consider the random walk  defined in  (\ref{model})-(\ref{poten}). There exist positive numbers
$\b_1$, $\b_2$ ($\b_1>\b_2$), $C_1$ and $C_2$  such that
\be\label{ballis}
\langle \o_N^2 \rangle\ge C_1N^2, ~~~~~~~~~~~~~~~~~~~~~~~\mbox{ for all $\b> \b_1$}
\ee
and
\be\label{diffu}
\langle \o_N^2 \rangle\le C_2N, ~~~~~~~~~~~~~~~~~~~~~~~\mbox{ for all $\b< \b_2$}
\ee
}

\vv

\\{\bf Remark}. Theorem above implies immediately that our random walk model
has two distinct regimes for the behavior of the end-to-end distance. Namely, inequality (\ref{ballis})
implies that the model defined
by (\ref{model})-(\ref{poten})
is ballistic (i.e. $\g=2$)
for all inverse temperature above $\b_1$ and  inequality (\ref{diffu})
implies that  the same  model is diffusive (i.e. $\g=1$)
for all inverse temperature below $\b_2$.

Observe that the theorem says
nothing about the behavior of the end-to-end distance
with $N$ at inverse temperature in the interval $\b\in [\b_2,\b_1]$. So, in principle, the existence of
an intermediate super-diffusive   phase for the present model in the region $\b\in [\b_2,\b_1]$
cannot be excluded.

\vv\vv\vv

\\\S2. {\it Proof of the theorem}
\vv
\\To prove the theorem above, following the ideas of \cite{bps}, we first define two further changes of variables.
Let first $\mu_i=\o_{i}-\o_{i-1}$, for $i=1,\dots ,N$.
Then $\mu_i\in\{\pm e_1,\pm e_2\}$, where $e_1=(1,0)$ and $e_2=(0,1)$. Moreover
$
\o_k=\sum_{i=1}^k \mu_i
$.
We now decompose the vectors $\mu_i$ in the following way
\be\label{deco}
\mu_i=\s_i \frac{e_1+e_2}{\sqrt 2}+\tilde\s_i \frac{e_1-e_2}{\sqrt 2}
\ee
with $\s_i,\tilde\s_i\in\{\pm 1\}$. We will denote a set of $\s_i,\tilde\s_i,\quad, 1\le i\le N$
as $\s_,\tilde\s$.
It is easy to see that the correspondence between a succession $\o\in\cal{W}_N$
and a set $\s_,\tilde\s$
is one-to-one and the probability measure $P(\o)$ is
mapped in
$$
P_\b(\o)=P_\b(\mu)=P_\b(\s,\tilde \s)= P_\b(\s)P_\b(\tilde \s)
$$
with
$$
P_\b(\s)= \frac{\exp\Bigg(+\b\sum\limits_{0\leq i<j\leq N}U^N_{ij}\s_i\s_j\Bigg)}{
\sum_{\s\in\Sigma_N}\exp\Bigg(+\b\sum_{0\leq i<j\leq N}U^N_{ij}\s_i\s_j\Bigg)
}
$$
and
\be \label{pot}
U^N_{ij}=\sum_{0\leq k\leq i<j\leq l\leq N} V_{kl}
\ee
Moreover observe that
$$
\o^2_N= \Big(\sum_{i=1}^N \m_i\Big)^2= \sum_{i=1}^n\sum_{j=1}^n \m_i\m_j= \sum_{i=1}^n\sum_{j=1}^n
(\s_i\s_j+\tilde\s_i\tilde\s_j)
$$

Hence we can rewrite (\ref{quadrado}) as

$$
\left\langle \o_N^2 \right\rangle=\sum_{\o\in\cal{W}_N}\o_N^2P_\b(\o)=
\sum_{\s\in\Sigma_N}\sum_{\tilde\s\in\Sigma_N}\sum_{0\leq i,j\leq N}(\s_i\s_j+\tilde\s_i\tilde\s_j)
P_\b(\s)P\_\b(\tilde\s)=
$$
\be\label{edss}
=2\sum_{0\leq i,j\leq N}\sum_{\s\in\Sigma_N}\s_i\s_j
P_\b(\s)
=2\sum_{0\leq i,j\leq N}\left\langle \s_i\s_j\right\rangle_{\cal N}^{\b U^N}
\ee

Therefore we have shown that the mean value of
the square end-to-end distance of the self-repelling random walk in $\Z^2$
is twice the mean value of $\sum_{0\leq i,j\leq N} \s_i\s_j$
in a one dimensional spin system on ${\cal N} = \{1,2,\dots ,N\}$ at inverse-temperature $\b$ with
free boundary conditions and ferromagnetic Hamiltonian given by
\be\label {Hs}
H^0_{\cal N}(\s_{\cal N})=-\sum_{1\le i<j\le N}U^N_{ij}\s_i\s_j
\ee
Note that the spin potential $U^N_{ij}$ defined
in (\ref{pot}) has some dependence on the volume $N$. However,
since $V_{ij}= |i-j|^{-\a}$, it is easy  to show
that for a fixed $\e>0$,  and all $i,j \in [\e N, (1-\e)N]$, there are constants $c_1,c_2$ depending on $\e$ such that
\be\label{bounds}
\bar U_{ij}^{(1)}=c_1 |i-j|^{-\a+2}\leq U^N_{ij}\leq c_2|i-j|^{-\a+2}=\bar U_{ij}^{(2)}
\ee
On the other hand, when $i,j$ are not both inside the region
${\cal N}_\e=\{ i\in {\cal N}: \,\,\,\e N\le i\le (1-\e)N \}$,
the potential
$U^N_{ij}$, that is evidently ferromagnetic
everywhere, can be bounded by
\be\label{bounds2}
0\leq U^N_{ij}\leq c_2|i-j|^{-\a+2}
\ee

Using now the Griffiths inequalities (valid for ferromegnetic systems) we have the following
bounds for the correlations.
\be\label{bounds3}
\left\langle \s_i\s_j\right\rangle_{{\cal N}_\e}^{\b\bar U^{(1)}}\leq
\left\langle \s_i\s_j\right\rangle_{\cal N}^{\b U^N}\leq
\left\langle \s_i\s_j\right\rangle_{{\cal N}}^{\b \bar U^{(2)}}
\ee

The inequality (\ref{bounds3}) is crucial for the purposes of this
note since it easily implies the diffusive-ballistic transition of
our model. It states that the quantity $\left\langle
\s_i\s_j\right\rangle_{\cal N}^{\b U^N}$, directly related to the mean
end-to-end distance via (\ref{edss}), is bounded above the by the
free boundary condition spin-spin correlation function of a
one-dimensional spin system in the volume $\cal
N$  with Hamiltonian given by
\be\label{H2}
H^2_{\cal N}=-\sum_{i<j\atop \{i,j\}\in {\cal N}}{c_2\over |i-j|^{\a-2}}\,\s_i\s_j
\ee
The inequality
(\ref{bounds3}) also states that the same quantity $\left\langle
\s_i\s_j\right\rangle_{\cal N}^{\b U^N}$ is bounded below by the free
boundary condition spin-spin (not truncated) correlation of an
analogous one-dimensional spin system, this
time in the smaller volume $\cal N_\e$  with  Hamiltonian given by
\be\label{H1}
H^1_{\cal N}=-\sum_{i<j\atop \{i,j\}\in {\cal N}}{c_1\over |i-j|^{\a-2}}\,\s_i\s_j
\ee

Now, it is a consequence of \cite{d} and \cite{fs} that, if $3<\a\le 4$, there exists a value $\b_1$ (depending on $c_1$) such that the spin system
with Hamiltonian (\ref{H1}) has spontaneous magnetization $m_*(\b)>0$  for all $\b>\b_1$. By \cite{l}, the infinite volume limit of the two point function
$\left\langle
\s_i\s_j\right\rangle_{{\cal N}_\e}^{\b\bar U^{(1)}}$ with free boundary condition
converges to $m_*^2$. In other words, for $N$ sufficiently large,
\be\label{b4}
\left\langle
\s_i\s_j\right\rangle_{{\cal N}_\e}^{\b\bar U^{(1)}}\ge {1\over 2}m_*^2(\b),\,\,\,\,\,\,\,\,\mbox{for all $\b>\b_1$}
\ee

Therefore, using the lower bounds (\ref{bounds3}) and (\ref{b4})and recalling that
$m(\b)$ is monotonic increasing with $\b$, we get,  for $N$ sufficiently large, and
for any  $\b>\b_1$
\be\label{result}
\left\langle \o_N^2 \right\rangle=2\sum_{0\leq i,j\leq N}\left\langle \s_i\s_j\right\rangle_{\cal N}^{\b U^N}
\ge C_1 N^2
\ee
where $C_1= (1-2\e)^2{m^2_*(\b_1)}$. This proves inequality (\ref{ballis}).



Concerning now the upper bound for $\left\langle \s_i\s_j\right\rangle_{{\cal N}}^{\b\bar U^{(2)}}$,
again we can use the classical results \cite{d} and \cite{fs} to claim that, if $3<\a\le 4$,
there exists an inverse temperature $\b_2$ (and $\b_2<\b_1$, since $c_2>c_1$) such that the spin system
with Hamiltonian (\ref{H2}), for all $\b<\b_2$, has no spontaneous magnetization (i.e. $m_*=0$)
and, by \cite{accn} (see also \cite{an} and \cite{sz}), there exists a positive constant $0<C_{\b_2}<+\infty$
such that the infinite volume limit of the two point function
$\left\langle
\s_i\s_j\right\rangle_{{\cal N}_\e}^{\b\bar U^{(2)}}$ with free boundary condition is bounded above
by $C_{\b_2}|i-j|^{-\a+2}$.
In other words, for $N$ sufficiently large
\be\label{b5}
\left\langle
\s_i\s_j\right\rangle_{{\cal N}_\e}^{\b\bar U^{(2)}}\le
{C_{\b_2}}{1\over|i-j|^{\a-2}},\,\,\,\,\,\,\,\mbox{for all $\b<\b_2$}
\ee

Thus, using the upper bounds (\ref{bounds3}) and (\ref{b5}), we get,  for $N$ sufficiently large, we get that
\be
\left\langle \o_N^2 \right\rangle=2\sum_{0\leq i,j\leq N}\left\langle \s_i\s_j\right\rangle_{\cal N}^{\b U^N}\le C_2N
\ee
\vv
\\whenever  $\b<\b_2$, with $C_2=2{C_{\b_2}}\sum_{i>0}i^{-\a +2}$.
This proves inequality (\ref{diffu}) and concludes the proof of the theorem.
\vv\vv\vv

\\\S3. {\it Concluding Remarks}.
\vv
\\As remarked above, the comparison method proposed in this note, based on the
inequalities (\ref{bounds})-(\ref{bounds3}), doesn't allow  to conclude
that  the random walk model defined by (\ref{model})-(\ref{poten}) exhibits a genuine diffusive-ballistic
phase transition with a unique critical point $\b_s$ inside the interval $[\b_1,\b_2]$ below which the system
is diffusive and above which the system is ballistic.
In order to prove that, one should  study directly the Ising model with Hamiltonian (\ref{Hs})
and show that results about the standard Ising model with
Hamiltonian (\ref{H2}) remain valid. In particular, one should prove that the two point correlation function is absolutely summable in the whole subcritical phase.
This task does not seem completely trivial since Hamiltonian (\ref{Hs}), differently from (\ref{H2}),
is not translational invariant.

The  theorem above  could be generalized  for random walks
described by (\ref{model})-(\ref{poten})
with any power  $\a\in \mathbb{R}^+$. As a matter of fact,
by the results contained in \cite{r},
it is immediate to conclude that our random walk model is diffusive at any temperature $\b>0$,
whenever $\a>4$. Moreover one also expects that the behavior of the present model is ballistic at  any temperature
for $\a\leq 2$. This  follows from the very reasonable claim that a one-dimensional
 spin system with ferromagnetic interaction proportional to $|i-j|^{-\l}$ with $\l<1$, should have
a (finite volume) non  zero magnetization for all values of the temperature $\beta>0$. However, as far as we know,
in the literature there is no analogous of the results contained in \cite{l} for one-dimensional spin systems
with non summable ferromagnetic interaction, since the infinite volume limit of such systems is not well defined
(see e.g. \cite{gm}).

Finally, the results obtained for this bidimensional model are clearly valid also in the case $d=1$. On the other hand,
a generalization  to dimensions $d>2$  is also possible but it appears technically more involved.
Namely, one would need to generalize
the works \cite{d} and \cite{fs} to a one-dimensional
spin system in which the spin variable $\s_i$ at the site $i$ takes value in the discrete set
$\{\pm e_1,\dots \pm e_d\}$ (the possible unit steps in a random walk in $\Z^d$)
with $e_i$ being the unit vector in $d$-dimensions parallel to the $i^{th}$ axis, by
proving  that also for such  spin systems there are two different
regimes with zero magnetization and non zero magnetization.

Note that, when $d\ge 3$ such a vector spin system does not admit anymore a simple decomposition into
independent Ising-type systems (i.e., with spin $\s=\pm1$) such as (\ref{deco}).
E.g., for $d=3$ there are only  6 possible states for the  spin at $i$ while 3 independent Ising model
have $2^3=8$ possible states.

\vv\vv
\\{\bf Acknowledgements}. AP was partially supported by CNPq and FAPEMIG. RS was partially supportd by
Pr\'o-Reitoria de Pesquisa-UFMG under grant 10023.
We thank Marzio Cassandro and Errico Presutti for useful discussion.

\end{document}